\documentclass[letterpaper,aps,pre,groupedaddress,superscriptaddress,floatfix,twocolumn,amsfonts,amssymb]{revtex4-1}

\usepackage{epsfig}
\usepackage{graphicx}
\usepackage{dcolumn}
\usepackage{bm}

\DeclareMathAlphabet{\mathpzc}{OT1}{pzc}{m}{it}


\newcommand{\beq}{\begin{equation}}
\newcommand{\eeq}{\end{equation}}
\newcommand{\bea}{\begin{eqnarray}}
\newcommand{\eea}{\end{eqnarray}}

\usepackage{xcolor}

\begin{document}

\title{Interatomic Potential in a Simple Dense Neural Network Representation}

\author{Ka-Ming Tam}
\affiliation {Department of Physics and Astronomy, Louisiana State University, Baton Rouge, LA70803}
\affiliation {Center for Computation and Technology, Louisiana State University, Baton Rouge, LA70803}

\author{Nicholas Walker}
\affiliation {Department of Physics and Astronomy, Louisiana State University, Baton Rouge, LA70803}

\author{Samuel Kellar}
\affiliation {Department of Physics and Astronomy, Louisiana State University, Baton Rouge, LA70803}

\author{Mark Jarrell}
\altaffiliation{Deceased 20th July, 2019}

\affiliation {Department of Physics and Astronomy, Louisiana State University, Baton Rouge, LA70803}
\affiliation {Center for Computation and Technology, Louisiana State University, Baton Rouge, LA70803}

\date{\today}

\begin{abstract}
Simulations at the atomic scale provide a direct and effective way to understand the mechanical properties of materials. In the regime of classical mechanics, simulations for the thermodynamic properties of metals and alloys can be done by either solving the equations of motion or performing Monte Carlo sampling. The key component for an accurate simulation of such physical systems to produce faithful physical quantities is the use of an appropriate potential or a force field. In this paper, we explore the use of methods from the realm of machine learning to overcome and bypass difficulties encountered when fitting potentials for atomic systems. Particularly, we will show that classical potentials can be represented by a dense neural network with good accuracy.
\end{abstract}

\maketitle

\section{Introduction}

The advance of computational chemistry and material science methods over the past few decades has resulted in the computation of rather accurate potential energy surfaces for many systems. However, the success of methods such as configuration interaction, coupled cluster theory, many body perturbation theory and the more recent density functional theory still face tremendous limitations with respect to the system sizes that can be simulated. While this may not pose a major problem for the study of small molecules and systems with perfectly periodic structure, this presents a crucial obstacle for the study of phenomena involving long length scales. Many mechanical properties, with notable research interest and significance, fall into this category, including defects, dislocations, and interfaces in which features related to length scales of more than just a several lattice constant play an important role, while the detailed structure of the nucleic and electronic orbitals are comparatively much less dominant in driving physical interactions. 

A viable approximation to bypass the computationally expensive quantum mechanical calculations of the nuclei and electrons for a macroscopic system is to parameterize the force or potential as a function of the positions of the nuclei alone. This strategy was pioneered with the Lennard-Jones potential about a century ago, where the two body potential between particles is composed of a repulsive core and an attractive tail which are both represented by inverse power terms of the distance separating the two bodies \cite{Jones_Chapman_1924a,Jones_Chapman_1924b}. The potential is expected to describe the interaction of the nobel gases rather well, but quantitatively accurate results are not expected for systems where covalent or ionic bonds dominate the interactions. Regardless of its shortcomings, the Lennard-Jones potential and its modifications are undoubtedly the most widely used potentials for different systems largely due to their simplicity. 

Clearly, more complex forms of potential than simple inverse power laws are required for the quantitative description of many physical properties. There are three major components to consider when fitting a new potential. These are comprised of the functional form (model) that the potential will take, the data that the potential is fit to, and the method used to perform the fit of the model to the data. 

For many simple physical systems, such as those composed of noble gases, a simple physical model can be readily applied. Unfortunately, in many circumstances, there is simply not a suitable a priori justification for such a simple model. Factors such as radial and angular displacements should be most relevant. For some cases, potentials describing only two-body interactions may not be able to properly described a physical system exhibiting many-body effects. For metallic systems, our main focus of the present study, electrons are de-localized, thus simple models of pair-wise interactions may not be sufficient to describe such systems. The most successful model which incorporates the idea of accounting for electronic structures described by density functional theory is the embedded atom model (EAM) \cite{Daw_Baskes_1984,Daw_Foiles_Baskes_1993,Ercolessi_Parrinello_Tosatti_1988}. The EAM consists of two components, the pair-wise potential, and the embedding function that accounts for the energy cost when an electron is put into the sea of de-localized electrons. A modified model that additionally accounts for the angular displacements of the atomic configurations compared to the EAM, dubbed as the Modified EAM \cite{Baskes_1992,Baskes_Johnson_1994}, has also been studied extensively. There have been many important papers devoted to the optimization of the EAM and MEAM potentials over the past couple of decades \cite{Jelinek_Groh_Horstemeyer_2012,Lee_Baskes_2000,Lee_Baskees_2001}. Other popular models for covalent systems such as carbon and silicon include Tersoff and Stillinger-Weber potentials \cite{Tersoff_1988,Stillinger_Weber_1985}.  

The paper is organized as follows. In the next section we provide a brief overview of the interatomic potential fitting. In the section III, we discuss using the dense neural network as the model for the interatomic potential. The detail of the method and the results are presented. We conclude and discuss a future development of utilizing the neural network for interatomic potential fitting in the last section. 

\section{Overview of Interatomic Potential Fitting}
\subsection{Reference States for Fitting}

Given a model or functional form for a potential, the most important choice for fitting the potential to measured data is the selection of targeted properties to fit to. Traditionally, the parameters in a model are fit to empirical properties such as lattice constants, cohesive energies, elastic constants, bulk moduli, and sometimes densities and melting temperatures \cite{Martinez_Yilmaz_Liang_etal_2013}. However, experimentally measurable quantities are rather limited and often come with non-negligible uncertainties. Nevertheless, if desire for a potential to faithfully reproduce particular empirical properties is present, the traditional approach may still be the best way to attain such a goal. Given the very limited availability of measurable quantities compared to the rather large set of parameters in more complex models which are sometimes tabulated continuous functions, a situation may arise where there is a multitude of viable fits for a particular model to the provided empirical data despite large variation in fitting parameters across the different fits. This is not just a simple inconvenience, as this leads to very practical problems in the transfer-ability of a given model fit. The obtained potential may fit very well to specific sets of empirical measurements for the specific phases that the empirical measurements were taken from, but the fit may not be accurate for quantities or phases outside the scope of the empirical measurements used for the fitting procedure. 

A remarkable development to improve the transfer-ability of potential fitting is to consider a much larger set of quantities to fit to -- a force matching approach \cite{Ercolessi_Adams_1994}. This approach becomes feasible with the development of large scale density functional theory (DFT) simulations. DFT provides more detail of the system at the atomic and electronic structure scales beyond the measured bulk physical quantities as well as calculations of potential energies and forces for each individual atom in the system. This vastly enlarges the features (or training sets) available for fitting the potential. Obviously, for such an approach to perform properly, the DFT data has to be accurate for not only the potential surface, but also ideally for the forces of each individual atom. The force matching approach allows much larger sets of training data which can be generated from DFT simulations for wide range of external parameters, such as temperature and pressure. Training data for surfaces, impurities, and interfaces can additionally be considered. Naively, due to the vastly expanded training data, different phases of the materials should be more properly accounted for. The hope is that such an approach should greatly enhance the transfer-ability of the potential fit. Essentially, the force matching method can be considered as a scheme of high dimensional interpolation among large number of reference configurations. Along this line of thought, traditional methods of fitting parameters to a small set of measurable physical quantities can be considered as a low dimensional extrapolation. After all, the quality of the fitted potential is still bounded by the assumption of the model. Most models should only presumably work within a limited range of external parameters. 

\subsection{Fitting Procedure}

Given the training data and the model, performing a potential fit is still a non-trivial problem, especially for the force-matching approach. Various techniques have been employed, such as conventional methods including the conjugate gradient and its variants.  More recently, genetic algorithms, and simulated annealing methods have also been explored. Broadly speaking, the fitting is composed of high dimensional non-linear optimization which is highly dependent on the model and the training data. In addition, the prescribed criterion for the goodness of fit, the penalty or cost function used for optimization, can affect the fitting \cite{Brommer_Gahler_2006,Brommer_Gahler_2007,Brommer_Kiselev_Schopf_etal_2015,Martinez_Yilmaz_Liang_etal_2013,Masia_Guardia_Nicolini_2014,Ercolessi_Adams_1994,Sun_etal_2018}. 

\section{Machine Learning Approach}

While many machine learning methods have been around for decades, explosive growth in use and interest has only been apparent in the last decade or so. This is largely driven by the availability of large data set and the need to extract information from said data in various academic and business sectors. This has lead to a rapid development both in terms of the development of the methods and the implementations of the methods. In particular, supervised machine learning approaches utilizing artificial neural networks have been applied extensively on solving many problems in science and engineering. The method itself is rather well studied and efficient implementations are easily accessible. The time is ripe to take advantage of these developments and apply them to atomic simulations. For example, research has been conducted on utilizing the machine learning approach in designing atomic potentials \cite{Chmiela2019,Behler2007,Bartok2010,Rupp2012,Botu2015,Li2015,Artrith2017,Glielmo2017,Zhang2018,Lubbers2018,Ryczko2018,Chmiela2017,Bartok_Kermode_Bernstein_2018,Zhang_Lin_Wang_etal_2019}. More recently, the unsupervised machine learning method has been used to detect phase transitions in metals \cite{Walker_Tam_Novak_2018}. 

A crucial step to employing the machine learning approach is to choose the appropriate input data. Ideally, the input data set should be invariant with respect to symmetries and of adequate size, but at the same time sufficient to describe the environment surrounding to the atom for which the potential or/and force is calculated \cite{Bartok2013,Zhang_etal_2018}. 
There are many proposals for the functional forms for describing the local environment which are invariant respect to different symmetries ( rotational, transnational, permutation). These include but not limited to Gaussian approximation potential \cite{Bartok_Kondor_Csanyi_2013,Dragoni_etal_2018}, generalized neural-network representation\cite{Behler_Parrinello_2007,Szlachta_Bartok_Csanyi_2014}, spectral neighbor analysis potential \cite{Thompson_etal_2015,Chen_etal_2017,Li_etal_2018}, moment tensor potential \cite{Shapeev_2016,Podryabinkin_Shapeev_2017} and orthogonal basis function approach \cite{Kocer_Mason_Erturk_2019}. 
We will show in the following section, a naive and non-invariant description of the local environment is capable to provide a rather good result.

\begin{figure}[bth]
\centerline{
\includegraphics*[height=0.31\textheight,width=0.5\textwidth, viewport=00 00 750 600,clip]{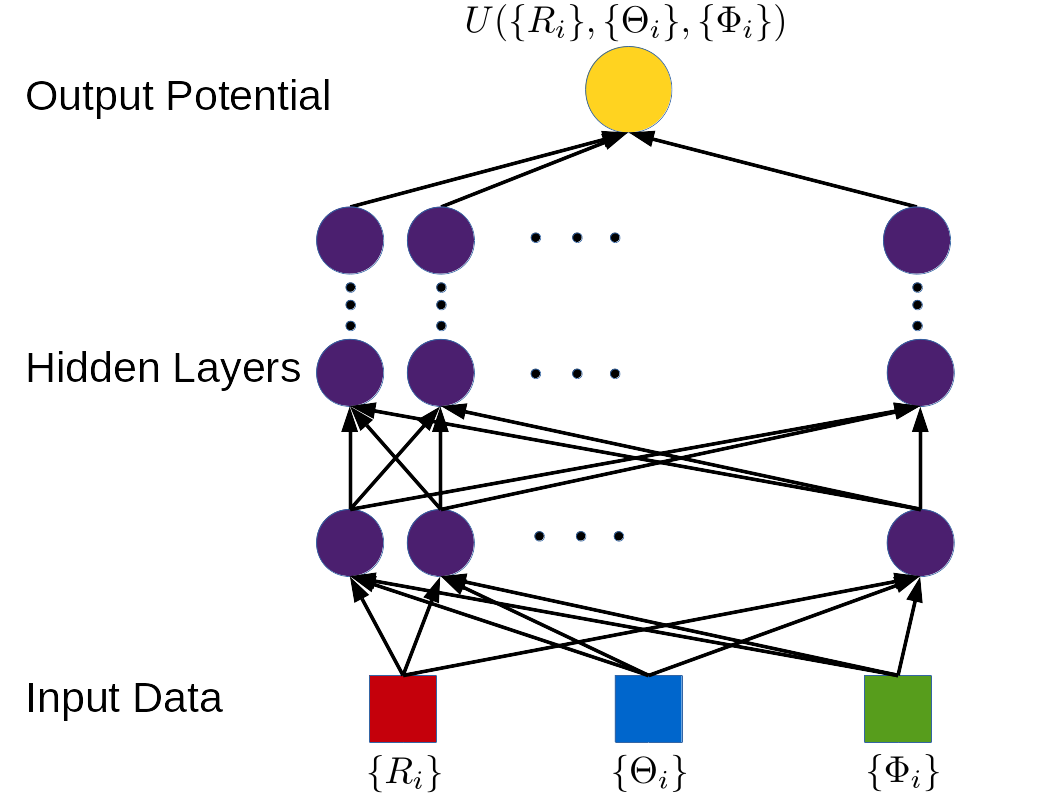} 
}
\caption{The potential is represented by a dense neural network with input, hidden, and output layers. Given a designated atom denoted as 0-th atom. The input data includes the radial distances $\lbrace R_{i} \rbrace$, polar angles $\lbrace \Theta_{i} \rbrace$, and azimuth angles $\lbrace \Phi_{i} \rbrace$ of $N_c$ number of atoms which are closest to the 0-th atom. Thus, the input data contains all the information about the location of the atoms in the vicinity of the 0-th atom for which the potential is calculated. See the section IIIB for the detail of the hidden layers. The output is the potential for the 0-th atom, $U$.}
\label{NN}
\end{figure}

The inherent limitation of fitting to a physical model is that a good model may not be known \textit{a priori}. Additionally, the external parameter range of fitting is also an important factor in deciding the quality of the fitted potential. For example, fitting the potential for high pressure conditions places more importance on the potential close to the nuclei; conversely, high temperature conditions place more importance on the tail of the potential. Most potentials are designed for crystal phase with periodic structure. The potential which is valid over multiple physical regimes is not known in most cases, or is perhaps rather complicated. Neural network potentials provide a new avenue for potential fitting which is perhaps more adaptive than conventional methods. 

Fitting to physical properties usually exhibits an under-fitting problem, in that many parameters are adjusted to only few fitting targets. In principle, a neural network allows for fitting to a large set of parameters. Specifically, a force matching approach can be employed in training the neural network. 

Another major difficulty of the potential fitting problem lies in identifying and approaching the best fit. The machine learning community has developed very efficient methods and implementations for fitting a neural network to training data. Therefore, one can mostly bypass the burden of devising a good method for fitting the parameters of the potentials.

\subsection{Neural network potential}

We devise a scheme to estimate the potential of any atom in a system composed of many atoms. First, the range of the potential is assumed to be finite. This assumption should be good for most metals and alloys in which the screening effects from the electron cloud can be justified. Second, for a generic potential, one cannot neglect the contribution from the effective many body coupling. For these reasons, the surrounding environment of a given atom can be described by the distance, the polar angle, and the azimuth angle with respect to the atom. A set of parameters ${R_1,R_2,...R_{N_c}},{\Theta_1,\Theta_2,...\Theta_{N_c}},{\Phi_1,\Phi_2,...\Phi_{N_c}}$ provide a complete description of the environment in the vicinity of an atom. For the practical reason of keeping a fixed input data size for the neural network input, instead of choosing a cutoff distance, we choose to use a cutoff number, $N_c$. For which the $N_{c}$ atoms which are closest to the atom in consideration are used as the input data. We sort the distance in ascending order, that is $R_{1} < R_{2} < ... < R_{N}$. 

We note that the above choice of describing the configuration in the vicinity of the atom of interest for calculating the potential is not unique. This choice is based on the fact that for atomic potentials, in particular for the description of metals and alloys, the distance plays a dominant role in deciding the potential. For this reason, we would like to emphasize this by explicitly inputting the distance as the training data. For simulations of molecules, the angle could play a dominant role, thus this may not always be the optimal choice.

\subsection{Training the neural network}

We demonstrate that a neural network is a very good candidate for generating a potential by fitting to the simulated configurations and corresponding potential from classical molecular dynamics. The simulations are done using the LAMMPS software package with two different potentials, EAM and MEAM, which are used for metals\cite{Sheng_Kramer_Cadien_2011,Pascuet_Fernandez_2015}. We use aluminum as an example, with the simulations starting from perfect FCC crystals with $5\times5\times5$ unit cells. 500 atoms are set at a random velocities corresponding to a temperature of 500K at zero pressure with the NPT ensemble. The pressures are fixed to zero and the temperature is increased at a constant rate from 500K to 2000K before decreasing back down at a constant rate to 500K over $30 \mu s$. One of the atoms is labelled as the target atom in which we are interested in its potential, we denote its as the $0$-th atom. We record the positions of the $54$ atoms which are closest to the $0$-th atom and the potential of the $0$-th atom is recorded. The distance, polar angles and azimuth angles of these $54$ atoms relative to the $0$-th atom together with the potential of the $0$-th atom are the training data for the neural network. The benchmark data is generated by heating up in a constant rate from 500K to 1000K in $6 \mu s$. $54$ atoms are included in the fitting as these are the total number of atoms up to the fourth nearest neighbors for a perfect FCC lattice.

\begin{figure}[bth]
\centerline{
\includegraphics*[height=0.36\textheight,width=0.50\textwidth, viewport=0 50 540 540,clip]{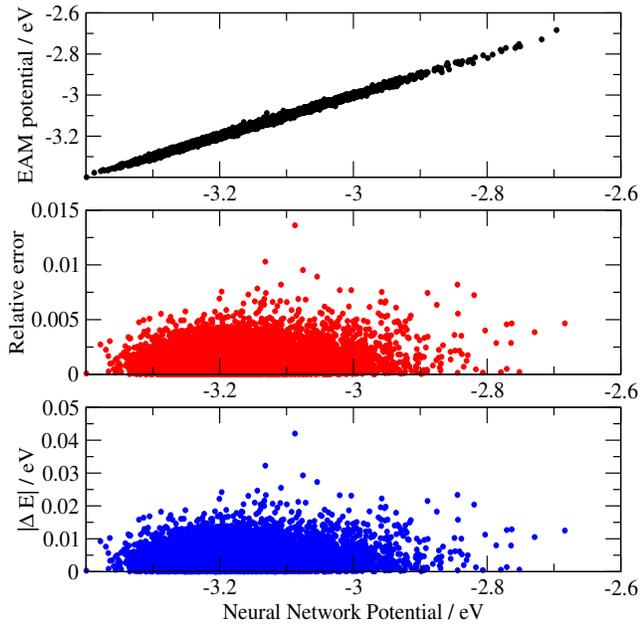} 
}
\caption{The neural network potential is trained with 60000 configurations. The configurations are generated by an EAM aluminum potential for a range of temperatures from 500K to 2000K at zero pressure \cite{Sheng_Kramer_Cadien_2011}. The upper figure shows the comparison of the neural network potential to the potential directly calculated by the EAM potential. A total of 10000 data points are shown in this benchmark. They are generated for temperatures between 500K and 1000K, where the range of temperatures around the the melting point are presumed to be difficult cases. A perfect match between the neural network potential and the EAM potential should exhibit linear behavior. Most points adhere rather closely to this target line. We further analyze the error of the neural network potential by calculating the percentage error relative to the EAM potential as shown in the middle figure. Except a small fraction of data points, most of them (over $77\%$) exhibit error less than $0.2\%$. The largest error is about $1.4\%$. The lower figure shows the absolute difference between the potentials in electronvolt.}
\label{eam}
\end{figure}

\begin{figure}[th]
\centerline{
\includegraphics*[height=0.36\textheight,width=0.50\textwidth, viewport=00 50 540 540,clip]{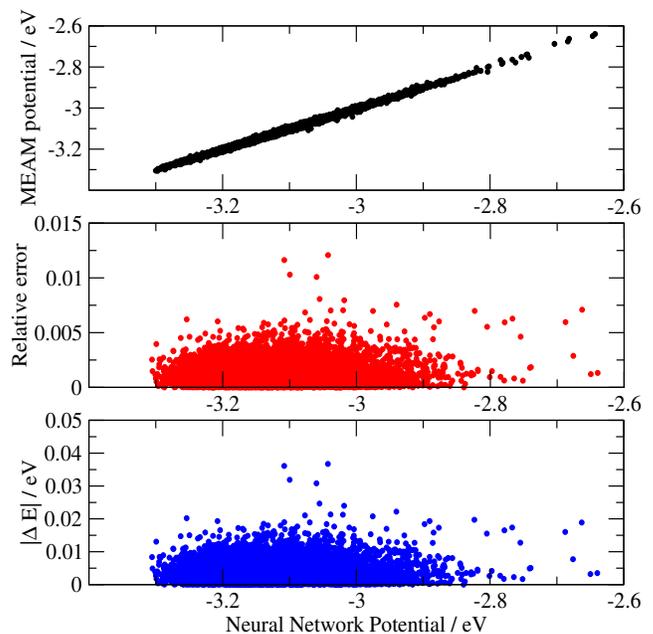} 
}
\caption{These figures are for the MEAM potential \cite{Pascuet_Fernandez_2015}. The data is generated with the same procedure as that for the EAM potential. See the caption of the fig. \ref{eam} for the details. Similar to that of the EAM potential, most points adhere rather closely to the target trend of the straight line. We further analyze the error of the neural network potential by calculating the percentage error relative to the MEAM potential as shown in the middle figure. Aside from a small fraction of data points, most of them (over $81\%$) exhibit error of less than $0.2\%$. The largest error is about $1.3\%$. The lower figure shows the absolute difference between the potentials in electronvolt.}
\label{meam}
\end{figure}

We are mostly interested in fitting the potential near the melting temperature, which is presumably the most difficult regime for attaining a good fitting. The training data should cover a wide range of temperatures, so as to consider substantially different configurations in the training to address the transfer-ability problem. If the crystal phase is the only region of interest, the training data can be simplified to include only configurations in the lower temperature range, which is essentially in the spirit of fitting to physically measurable parameter within a narrow range of temperatures. One of the goals of this neural network approach is to widen the transfer-ability of the potential, therefore we also consider the range of temperature much higher than the melting point. The benchmark data is set around the melting point as a litmus test for the quality of the fit. 

The neural network we consider is a dense neural network, as seen in Fig. \ref{NN}. There are $3\times54=162$ input variables, which include ${R_1,R_2, ...R_{54}},{\Theta_1,\Theta_2, ...\Theta_{54}},{\Phi_1,\Phi_2, ...\Phi_{54}}$ against the target atom at the $0$-th index. There are 9 dense hidden layers consisting of 162, 81, 81, 42, 42, 21, 21, 11, and 11 units with the output unit corresponding to the potential of the target atom. See Fig. \ref{NN} for more detail. We train the neural network on 60000 training sample configurations from the simulations over 1000 iterations. It takes less than an hour on an Intel i7-4790 desktop using the Keras with Tensorflow as the back-end \cite{chollet2015keras,tensorflow2015-whitepaper}. 

The results are shown in the Fig. \ref{eam} and \ref{meam} for the EAM and MEAM potentials respectively.
The comparison between the neural network generated potential and the benchmark potential are shown in the upper panels. A perfect fit would have all data points to fall on a straight line. The quality of the fit is rather good and almost all points adhere to the straight line for both the EAM and MEAM potential. We analyze the error, by plotting the relative error as a function of potential in the middle panels and the absolute difference in electronvolt in the lower panels. We find that most of the data exhibits relative error of less than $0.2\%$. There are few data points which have visibly larger relative error, but none of them are beyond $1.4\%$.

\begin{figure}[bth]
\centerline{
\includegraphics*[height=0.26\textheight,width=0.52\textwidth, viewport=00 50 720 540,clip]{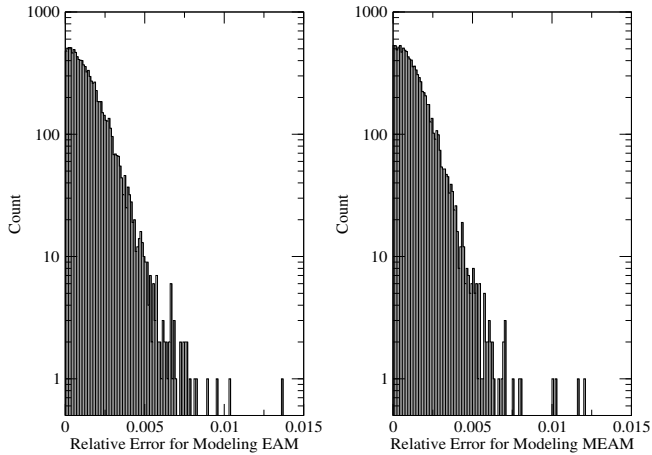} 
}
\caption{The distributions of the relative errors from the neural network potential for both EAM and MEAM potentials. Each histogram is constructed with 10000 points in 150 uniform bins of width $= 0.0001$.}
\label{error_dis}
\end{figure}

We further analyze the error by plotting the distributions of the relative error for both potentials in Fig. \ref{error_dis}.
It is noteworthy that the quality of the fit judged by the distribution of the errors for the two different potentials does show some difference. The fit to the MEAM potential, however, does not have substantial deterioration compared to that of the EAM potential. This difference notwithstanding, the errors are rather small for both cases. This demonstrates that the neural network is capable of providing a good fit with no severe sensitivity to the complexity of the potential. This is an encouraging signal that the neural network potential is rather adaptive. 

The difference between the MEAM potential and the EAM potential is in the angular dependence of the electrons charge density term. The EAM potential does not have explicit angular dependence, it renders the input of angular dependence as redundant. In general, we do not have a priori knowledge about the presence or absence of angular dependence, thus we include it in the training data set. 

\begin{figure}[bth]
\centerline{
\includegraphics*[height=0.26\textheight,width=0.52\textwidth, viewport=00 50 720 540,clip]{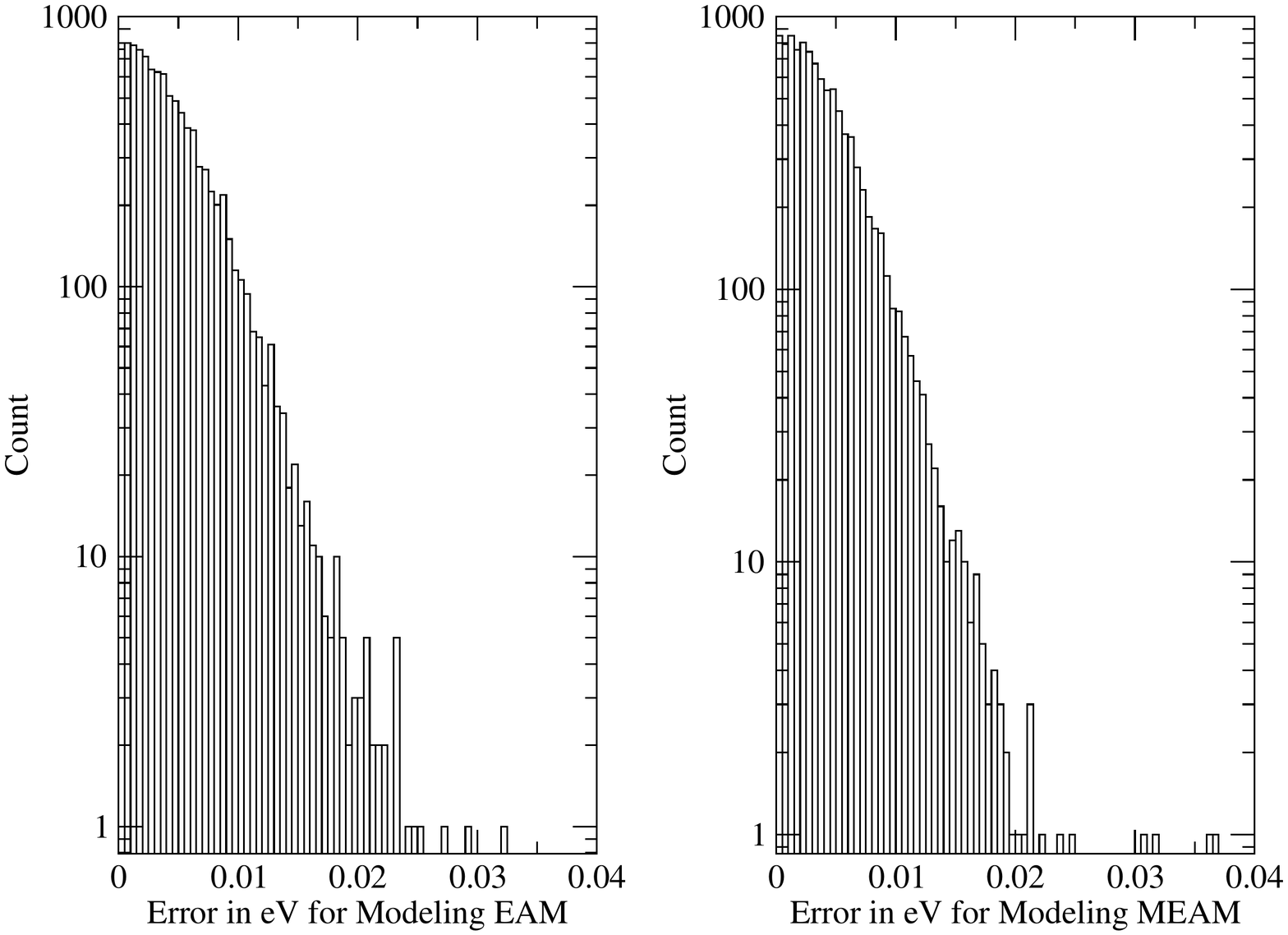} 
}
\caption{The distributions of the absolute error in energy from the neural network potential for both EAM and MEAM potentials. Each histogram is constructed with 10000 points in 50 uniform bins of width $= 0.001$eV.}
\label{error_energy_dis}
\end{figure}

We further analyze the errors of the neural network potentials by plotting the distribution of the absolute energy difference in the fig. \ref{error_energy_dis}. For fitting into the EAM potential, more than $66\%$ samples have less than 5meV error and more than $93\%$ samples have less than 10meV error. For fitting into the MEAM potential, more than $71\%$ samples have less than 5meV error and more than $95\%$ samples have less than 10meV error. 

\section{Summary and Outlook}

We discuss the three major challenges of potential fitting in the section I and II, they are the transfer-ability of the model, flexibility of the model, and the optimization of the model. A neural network potential provides a new avenue for tackling these problems. For the problem of transfer-ability, the machine learning approach is naturally designed to adapt to a large pool of vastly different samples in data set. Together with the force-matching approach, it can be used to fit the data from the low temperature crystal phase up to beyond the melting temperature. This should allow the potential to be well adapted to many cases. We demonstrate this point by benchmarking the neural network potential across the melting point. 

For the problem of flexibility of the model, we have demonstrated that by fitting a neural network to data generated with two different widely used potentials, the qualities of fits are rather similar, despite considerable differences in the potentials themselves. This shows the neural network is flexible in adapting to different models. Indeed, it has been shown recently that a neural network can be used to fit to the wavefunction of quantum systems \cite{Carleo_Troyer_2016}. With this in mind, we believe a simple dense neural network as that in this paper can be used to fit the potential surfaces from quantum mechanical \textit{ab initio} calculations. We note that obtaining accurate potential energy surface from the \textit{ab  initio} poses a major challenge, especially when the system is not in a crystal phase with periodic structure.

For the difficulty in optimizing the model, we basically take advantage of the developments made in the machine learning community to produce robust neural network architectures. Thus, there is no necessity to investigate a new method for optimizing the neural network. As a result, the present approach tackles the three major issues in potential fitting process discussed above. 

Whether a neural network potential can generate and predict the physical properties of materials is highly dependent on the quality of the training data provided. It seems that the ``best'' potentials are often not solely based on the ab initio data. A combination of experimental data and ab initio data may provide better agreement with experiments, which can easily be accounted for using the framework described in this work\cite{Lee_Baskees_2001}. The method discusses in this paper provides a simple but rather accurate method for obtaining inter-atomic potential by utilizing the development of the neural network research. Even within the present approach, the fitting can be further improved by fine tuning the parameters in training the neural network. Additionally, a possible improvement is to employ methods, such as data augmentation, which can preserve or approximate the rotational invariance of the potentials \cite{Dieleman_2016,Macros_2016,Dieleman_2015,Quiroga_2019}. 

\section{Acknowledgement}
This work is funded by the NSF EPSCoR CIMM project under award OIA-1541079.
Additional support (MJ) was provided by the DOE Materials Theory grant DE-SC0017861.
Portions of this research were conducted with high performance computing resources provided by Louisiana State University (http://www.hpc.lsu.edu). An award of computer time was provided by the INCITE program. This research also used resources of the Oak Ridge Leadership Computing Facility, which is a DOE Office of Science User Facility supported under Contract DE-AC05-00OR22725.
\bibliographystyle{apsrev4-1}
\bibliography{refs}

\end{document}